\documentclass[conference]{IEEEtran}
\IEEEoverridecommandlockouts
\usepackage{amsmath,amsfonts,amssymb}
\usepackage{algorithmic}
\usepackage{algorithm}
\usepackage{array}
\usepackage[caption=false,font=normalsize,labelfont=sf,textfont=sf]{subfig}
\usepackage{textcomp}
\usepackage{stfloats}
\usepackage{url}
\usepackage{tabularx}
\usepackage{array}
\usepackage{svg}
\usepackage{booktabs}
\usepackage{verbatim}
\usepackage{graphicx}
\usepackage{mathtools}
\usepackage{cite}
\usepackage{xcolor}
\usepackage{stfloats}
\usepackage{enumitem}
\usepackage{makecell}
\newcolumntype{C}[1]{>{\centering\arraybackslash}m{#1}}
\def\BibTeX{{\rm B\kern-.05em{\sc i\kern-.025em b}\kern-.08em
    T\kern-.1667em\lower.7ex\hbox{E}\kern-.125emX}}
\begin{document}

\title{Multi-Turn Distributed Inference with \\ Mixture of Experts for 6G Edge--Cloud Networks}

\author{
\IEEEauthorblockN{Bo Liu, Haiyuan Li, Yuelin Liu, Yulei Wu, Rasheed Hussain, Shadi Moazzeni, Dimitra Simeonidou}
\IEEEauthorblockA{
\textit{Smart Internet Lab,} 
\textit{School of Electrical, Electronic and Mechanical Engineering (EEME), University of Bristol,} 
\textit{U.K.} \\
E-mail: \{bo.liu, ocean.h.li, y.l.wu, yuelin.liu, rasheed.hussain, shadi.moazzeni, dimitra.simeonidou\}@bristol.ac.uk
}
}

\maketitle
\begin{abstract}

Mixture-of-Experts (MoE) architectures are increasingly deployed across 6G edge--cloud networks, where sparse activation reduces the computational footprint of each inference to only a fraction of the full expert set. However, MoE inference in edge-cloud networks creates a tension between KV state locality and elastic expert dispatch. KV state relocation incurs substantial transfer overhead, while expert computation benefits from spreading across the network to exploit available capacity. This tension is amplified in multi-turn inference, where each turn extends the KV state that must persist across the dialogue.
To this end, we present StateFlow, a distributed inference policy that decouples persistent KV state from transient sparse computation. StateFlow pins KV state at a sticky serving site for cross-turn reuse and jointly optimizes expert dispatch and aggregation placement across the network.
We further implement a real-world testbed with kernel-level network emulation and experimental results show that StateFlow sustains more than $2\times$ higher stable dialogue concurrency than the distributed baseline solutions and reduces turn-level p95 latency under multi-turn inference by 53.0\%.
\end{abstract}


% \vspace{0.3cm}
\section{Introduction}
The emergence of 6G edge-cloud networks, spanning access edges, edge clouds, and remote clouds, creates new opportunities for distributing generative-AI inference closer to end users \cite{letaief2021edge,duan2022distributed}. Mixture-of-Experts (MoE) architectures take this further by replacing the feed-forward component of each transformer layer with a set of parallel expert subnetworks, of which only a small subset is activated per token through sparse routing. This reduces the computational footprint of each inference and allows the full expert set to be distributed across the network. As experts reside on different network sites, each inference exchanges activations with the selected experts and gathers outputs back at every transformer layer.

This cross-site execution introduces a tension between two competing requirements. The Key-Value (KV) cache produced during inference must reside at a stable location, as relocating it incurs substantial transfer overhead proportional to the accumulated context length. Meanwhile, expert computation benefits from elastic dispatch across the network to exploit available capacity beyond the hosting site. 

For dialogue-oriented workloads such as conversational assistants, copilots, and agents, this tension is compounded across multiple \emph{turns}, each corresponding to one user query and model response that extends the shared KV state.
Successive turns within the same dialogue share an ever-growing KV state that must persist across the full dialogue lifecycle. Treating each turn as an independent request risks repeated dialogue migration, KV state reconstruction, and redundant cross-site transfers, all consuming the limited bandwidth and latency budget of the edge--cloud network.

In response, existing work approaches this problem from three directions. Collaborative edge inference \cite{kang2017neurosurgeon,teerapittayanon2017distributed,hu2019dynamic} partitions dense models across device-edge-cloud tiers but targets stateless pipelines without persistent dialogue state. LLM serving systems \cite{kwon2023efficient,patel2024splitwise,zhong2024distserve,sun2024llumnix} improve KV-cache management and multi-turn reuse \cite{gao2024cost,yu2025stateful}, but assume centralized execution and do not coordinate with distributed expert dispatch. Communication-aware MoE schemes \cite{xue2025wdmoe,kong2025serving,yang2025quality} optimize expert routing under network constraints, but operate at request granularity without accounting for the dialogue state accumulated over turns. Across all three directions, persistent session state and transient sparse computation have not been jointly orchestrated over edge-cloud networks.

Motivated by these gaps, we propose \textit{StateFlow}, a distributed multi-turn inference policy that decouples persistent KV state from transient sparse computation over 6G edge-cloud networks. StateFlow anchors dialogue ownership and KV state at a serving site to preserve cross-turn continuity, and dispatches sparse expert branches across the network to exploit available capacity. The main contributions are as follows:

\begin{itemize}[leftmargin=*]
\item \textit{Multi-Turn inference model:}
We formulate multi-turn distributed MoE inference as a stateful orchestration problem, in which persistent KV state and transient expert computation are treated as separate objects with distinct locality and elasticity requirements. This formulation departs from per-request offloading models that implicitly couple state and computation into a single migration unit.

\item \textit{StateFlow execution policy:} Based on this formulation, we propose StateFlow, which preserves dialogue state through sticky owner selection and jointly optimizes latency-aware expert dispatch, path-aware aggregation placement, and online congestion adaptation across the network.

\item \textit{Real system evaluation:} To validate StateFlow under realistic network conditions, we implement it on a real-world testbed with kernel-level network emulation. Experiments show that StateFlow sustains more than $2\times$ higher stable dialogue concurrency than all baselines and reduces turn-level p95 completion latency by 53.0\%.
\end{itemize}

\section{Related Work}
\label{sec:related}

\emph{1) Collaborative Edge Inference}: Early efforts to push neural inference toward the edge focus on partitioning dense models across the device--edge--cloud hierarchy. Neurosurgeon\cite{kang2017neurosurgeon} profiles per-layer compute and communication cost to identify an optimal split point between a mobile device and a cloud server. DDNN\cite{teerapittayanon2017distributed} extends this idea to a three-tier setup that distributes layers across end devices, edges, and clouds. DADS\cite{hu2019dynamic} formulates partitioning as a constrained optimization and adapts the split point as bandwidth and load conditions evolve. These systems establish that cross-tier model partitioning is a viable performance lever, but they target dense feed forward models and stateless inference pipelines.

\emph{2) LLM Serving Systems}: vLLM\cite{kwon2023efficient} introduces PagedAttention, which manages the KV cache as paged virtual memory to eliminate fragmentation. Splitwise\cite{patel2024splitwise} and DistServe\cite{zhong2024distserve} disaggregate prefill and decode onto separate hardware so that the compute-bound and memory-bound phases no longer interfere. Llumnix\cite{sun2024llumnix} adds a runtime scheduler that performs live request migration to rebalance load. HexGen\cite{jiang2023hexgen} schedules parallel execution across devices with mixed accelerators, and EdgeShard\cite{zhang2024edgeshard} shards parameters across memory-constrained edge nodes. For multi-turn dialogue, CachedAttention\cite{gao2024cost} and Pensieve\cite{yu2025stateful} reuse KV state across turns to avoid redundant prefill. These systems largely assume homogeneous datacenter networks, and have yet to coordinate persistent state with sparse expert computation distributed across sites.

\emph{3) Sparse MoE Serving}: Switch Transformer\cite{fedus2022switch} establishes per-token sparse expert activation, scaling parameters without proportionally scaling compute. Building on this foundation, WDMoE\cite{xue2025wdmoe} studies wireless distributed MoE where experts reside on different edge nodes, jointly optimizing routing weights and channel conditions. Kong et al.\cite{kong2025serving} propose dynamic expert swapping on edge devices, preloading likely experts and evicting cold ones. Yang et al.\cite{yang2025quality} consider QoS-aware routing across expert servers based on per-server queue and latency state. These works recognize communication as a first-class concern for distributed MoE, but remain scoped to a single request and do not consider the interaction between persistent KV state and expert dispatch over many turns.

\section{System Model and Problem Formulation}
\label{sec:system_model}

We consider an edge--cloud network $\mathcal{N}=(\mathcal{V},\mathcal{L})$, where $\mathcal{V}$ is partitioned into three tiers:
access-edge ($\mathcal{V}_1$), edge-cloud ($\mathcal{V}_2$), and cloud-core ($\mathcal{V}_3$). Each site $v\in\mathcal{V}$ has different computation capacities $C_v(t)$, memory budget $M_v(t)$, and utilization state $u_v(t)$. Each link $(u,v)\in\mathcal{L}$ is characterized by propagation delay
$d_{uv}$ and available bandwidth $b_{uv}(t)$.
The inference workload targets multi-turn dialogues built on MoE models. In each MoE layer $\ell\in\{1,\dots,L\}$, the dense feed forward network is replaced by a set of experts $\mathcal{M}_{\ell}$ and a lightweight gating network. Experts are deployed across the network.
For each layer $\ell$ and expert $m\in\mathcal{M}_{\ell}$,
let $\nu_{\ell}(m)\in\mathcal{V}$
denote the site hosting expert $m$ of layer $\ell$.

The system serves a set of dialogues $\mathcal{S}$. Each dialogue $s\in\mathcal{S}$ evolves over turns $\tau=1,2,\ldots$.
The request first reaches an entry site $e_s\in\mathcal{V}_1$, and the inference process is then anchored at an owner site $o_s\in\mathcal{V}$, whose primary role is to maintain the persistent KV state across turns.
The owner is kept stable across turns whenever possible,  so that subsequent turns can reuse the anchored KV state without incurring unnecessary cross-site state migration.

For dialogue $s$ and transformer layer $\ell$, the owner site maintains the KV state
\begin{equation}
\mathbf{K}_{s,\ell}^{(\tau)},\;\mathbf{V}_{s,\ell}^{(\tau)}
\in \mathbb{R}^{H \times T_{s,\tau} \times d_h},
\label{eq:KV}
\end{equation}
where $H$ is the number of KV heads, $d_h$ is the per-head dimension,
and $T_{s,\tau}$ is the cumulative context length. 
The KV state is pinned at the owner. Each new turn contributes $n_{s,\tau}$ tokens, so the context grows as
\begin{equation}
T_{s,\tau}
=
T_{s,\tau-1}+n_{s,\tau},
\qquad
T_{s,0}=0.
\label{eq:ctx_growth}
\end{equation}

At turn $\tau$ and layer $\ell$, the hidden state for the new tokens is generated by the attention sub-layer using the cached context:
\begin{equation}
\mathbf{h}_{s,\tau,\ell}
=
\mathrm{Attn}_{\ell}\!\bigl(
x_{s,\tau},
\mathbf{K}_{s,\ell}^{(\tau-1)},
\mathbf{V}_{s,\ell}^{(\tau-1)}
\bigr),
\label{eq:owner_hidden}
\end{equation}
where $x_{s,\tau}$ is the input
embedding of the new tokens.

For a payload of size $p$ transmitted from site $u$ to site $v$, we define the end-to-end transfer latency as:
\begin{equation}
\bar{T}_{uv}(p,t)
=
d_{uv}
+
\frac{p}{b_{uv}(t)}
\label{eq:transfer_latency_moe}
\end{equation}
For an activated expert $m$ at layer $\ell$, its contribution to the
layer latency consists of three sequential phases: dispatching the
hidden activations from the owner site to the site hosting the selected
expert, executing the expert computation at that site, and returning the
expert output to the aggregation site. These three phase latencies are
defined as
\begin{align}
T^{\mathrm{d}}_{s,\tau,\ell,m}
&=
\bar{T}_{o_s,v_m}\!\left(\xi_{s,\tau,\ell},\,t\right),
\label{eq:dispatch_latency_moe}\\
T^{\mathrm{c}}_{s,\tau,\ell,m}
&=
\frac{w_{\ell,m}}{\tilde{C}_{v_m}(t)},
\label{eq:expert_exec_latency}\\
T^{\mathrm{r}}_{s,\tau,\ell,m}(a)
&=
\bar{T}_{v_m,\,a}\!\left(\rho_{s,\tau,\ell,m},\,t\right),
\label{eq:return_latency_moe}
\end{align}
where $\xi_{s,\tau,\ell}$ denotes the size of the dispatched activations, $w_{\ell,m}$ is the workload of expert $m$ at layer $\ell$, $\tilde{C}_{v_m}(t)$ represents the effective service rate at site $v_m$, and $\rho_{s,\tau,\ell,m}$ denotes the size of the expert output.

Based on the above three phase latencies decomposition, the latency of 
MoE layer $\ell$ under the aggregation site $a$ is
\begin{align}
\Lambda_{s,\tau,\ell}(a) = & \max_{m \in \mathcal{M}^{\star}_{s,\tau,\ell}}
\Big[ T^{\mathrm{d}}_{s,\tau,\ell,m} + T^{\mathrm{c}}_{s,\tau,\ell,m} 
+ T^{\mathrm{r}}_{s,\tau,\ell,m}(a) \Big] \nonumber \\
& + \mathbf{1}[a \neq o_s] \cdot \bar{T}_{a,o_s}(\eta_{s,\tau,\ell}, t).
\end{align}
where $\eta_{s,\tau,\ell}$ is the size of the merged output, and 
$\mathbf{1}[a \neq o_s]$ equals $1$ when aggregation occurs at a remote expert site and $0$ when aggregation already happens at the owner site, capturing the continuity cost of returning the merged result to the owner.

The latency of turn $\tau$ is obtained by accumulating the
latencies of all MoE layers:
\begin{equation}
D_{s,\tau}
=
\sum_{\ell=1}^{L}
\Lambda_{s,\tau,\ell}\!\left(a_{s,\tau,\ell}\right).
\label{eq:turn_latency_moe}
\end{equation}
Here, $D_{s,\tau}$ denotes the end-to-end completion latency
of one inference turn of dialogue $s$, measured from the
arrival of the turn request to the completion of the generated
response.

For an offered concurrency level $W$, let $\mathcal{S}(W)$
denote the set of admitted dialogues. Service success is
evaluated over the whole dialogue lifecycle. A dialogue is
counted as successfully served only if all turns satisfy
the turn-level latency budget:
\begin{equation}
\chi_s = \mathbf{1}\!\left[\,\max_{1 \leq \tau \leq \mathcal{T}_s} D_{s,\tau} \leq \Delta^{\mathrm{turn}}\right],
\end{equation}
where $\mathcal{T}_s$ is the number of turns in dialogue $s$, and $\Delta^{\mathrm{turn}}$ is the end-to-end turn completion  latency budget.

The resulting goodput is defined as the number of dialogues that complete their full multi-turn execution without
violating the turn-level latency budget:

\begin{equation}
\mathcal{G}(W)
=
\sum_{s\in\mathcal{S}(W)}
\chi_s .
\label{eq:goodput_moe}
\end{equation}

The objective of this paper is to maximize \(\mathcal{G}(W)\) by jointly optimizing owner selection, expert activation, and aggregation decisions.

\begin{subequations}
\label{eq:p1_moe}
\vspace{-0.4cm}
\begin{alignat}{2}
\mathbf{P1:}\quad
& \max_{\substack{\{o_s\},\,\{z_{s,\tau,\ell,m}\},\\ \{a_{s,\tau,\ell}\}}}
\quad \mathcal{G}(W)
\label{eq:p1_obj_moe}
\\[0.5mm]
\text{s.t.}\quad
& \max_{1\le \tau \le \mathcal{T}_s} D_{s,\tau} \le \Delta^{\mathrm{turn}},
&\qquad& \forall s,
\label{eq:p1_slo_moe}
\\
& \operatorname{loc}\!\left(K_{s,\tau}\right)=o_s,
&\qquad& \forall s,\tau,
\label{eq:p1_state_moe}
\\
& \sum_{\mathclap{m\in\mathcal{M}_{\ell}}} z_{s,\tau,\ell,m}=K_{\mathrm{top}},
&\qquad& \forall s,\tau,\ell,
\label{eq:p1_topk_moe}
\\
& \sum_{\mathclap{u\in\mathcal{U}_v(t)}} \kappa_u(t)\le C_v(t),
&\qquad& \forall v,t,
\label{eq:p1_compute_moe}
\\
& \sum_{\mathclap{u\in\mathcal{U}_v(t)}} \mu_u(t)\le M_v(t),
&\qquad& \forall v,t,
\label{eq:p1_memory_moe}
\\
& \sum_{\mathclap{f\in\mathcal{F}_{u,v}(t)}} \phi_f(t)\le b_{u,v}(t)\Delta t,
&\qquad& \forall (u,v),t.
\label{eq:p1_link_moe}
\end{alignat}
\label{all_eq}
\end{subequations}
Here, constraint~\eqref{eq:p1_slo_moe} enforces the per-turn latency 
budget across the dialogue lifecycle. Constraint~\eqref{eq:p1_state_moe} 
pins the KV state of dialogue $s$ at its owner site $o_s$. 
Constraint~\eqref{eq:p1_topk_moe} reflects the sparse activation pattern of MoE inference, where $z_{s,\tau,\ell,m} \in \{0,1\}$ is the binary expert activation indicator that equals $1$ if expert $m$ at layer $\ell$ is selected for turn $\tau$ of dialogue $s$ and $0$ otherwise. 
Constraints~\eqref{eq:p1_compute_moe} and~\eqref{eq:p1_memory_moe} 
respectively bound the aggregate compute and memory demand of all 
active computation units $\mathcal{U}_v(t)$ at site $v$, with 
per-unit demands $\kappa_u(t)$ and $\mu_u(t)$. 
Constraint~\eqref{eq:p1_link_moe} ensures that the total volume of 
active cross-site expert flows $\mathcal{F}_{u,v}(t)$ on link $(u,v)$ 
within an interval $\Delta t$ does not exceed the available link 
capacity.

\begin{figure}[!t]
    \centering
    \includegraphics[
        width=\columnwidth,
        trim=190 170 390 80, % 左 下 右 上
        clip
    ]{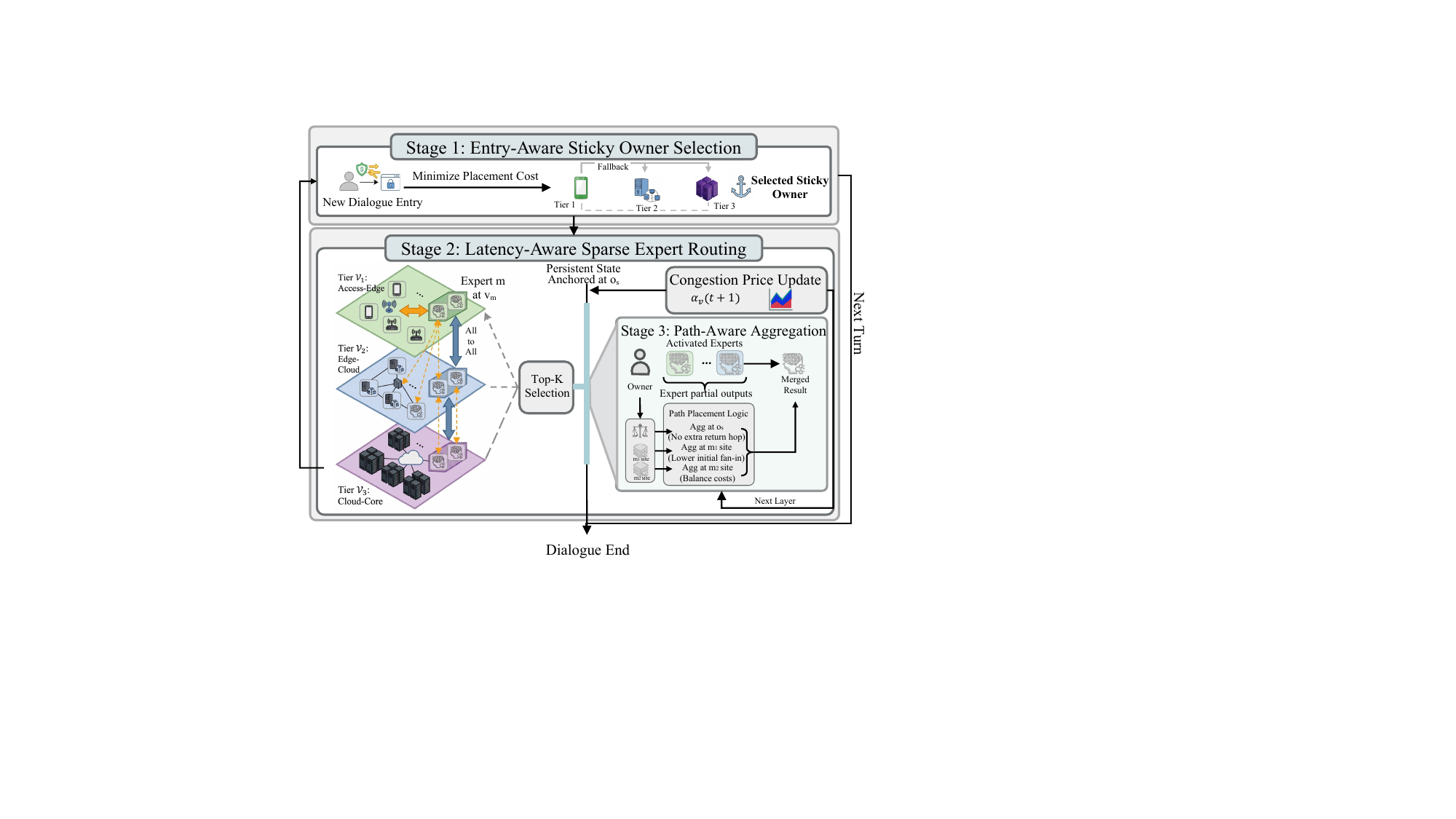}
    \vspace{-0.8cm}
    \caption{StateFlow execution policy: owner selection and per-layer expert routing with path-aware aggregation.}
    \vspace{-0cm}
    \label{fig:stateflow_policy}
    \vspace{-2mm}
\end{figure}

\section{StateFlow Execution Policy}
\label{sec:policy}

Directly solving problem~\eqref{all_eq} is intractable under dynamic arrivals, time-varying link conditions, and combinatorial sparse routing decisions. We develop \textit{StateFlow} as an online approximation policy. StateFlow optimizes a decomposed latency and congestion whose decisions map to owner selection, sparse expert activation, and aggregation placement. The policy preserves persistent state through sticky owner selection, while adapting sparse computation and aggregation online to current network and load conditions, as illustrated in Fig.~\ref{fig:stateflow_policy}. The policy includes entry-aware sticky owner selection, latency-aware sparse expert routing, and path-aware aggregation.

%\vspace{-0.1cm}
\subsection{Entry-Aware Sticky Owner Selection}
%\vspace{-0.1cm}
For a new dialogue $s$, the gateway determines the entry site $e_s\in\mathcal{V}_1$, then selects the owner by minimizing a weighted placement cost over a candidate set:
\begin{equation}
o_s = \arg\min_{v \in \mathcal{P}(e_s)} 
\Big[ \omega_1 \bar{T}_{e_s, v}(\xi_0, t) + \omega_2 u_v(t) \Big],
\label{eq:owner_select}
\end{equation}
where $\mathcal{P}(e_s)\subseteq\mathcal{V}_1$ denotes the set of peer sites available as fallback options, $\xi_0$ is the initial state payload size, and $u_v(t)$ is the normalized utilization of site $v$.

Once selected, the owner remains sticky whenever possible, so that the dialogue's persistent KV state is preserved at a stable edge site across turns. This design avoids repeated KV state migration as the dialogue evolves, thereby reducing the inter-site transfer overhead required to maintain continuity.

%\vspace{-0.1cm}
\subsection{Latency-Aware Sparse Expert Routing}
%\vspace{-0.1cm}

At each turn $\tau$ and layer $\ell$, the owner computes local attention and routing, then scores each expert $m\in\mathcal{M}_{\ell}$ using
\begin{equation}
\begin{aligned}
S_{s,\tau,\ell,m}
={}&\lambda\,R_{s,\tau,\ell}(m \mid h_{s,\tau,\ell})
-\beta\,T^{\mathrm{d}}_{s,\tau,\ell,m} \\
&-\gamma\,T^{\mathrm{c}}_{s,\tau,\ell,m}
-\alpha_{v_m}(t),
\end{aligned}
\label{eq:expert_score}
\end{equation}
where $R_{s,\tau,\ell}(m \mid h_{s,\tau,\ell})$ is the router relevance score and $\alpha_{v_m}(t)$ is the congestion price. The activated expert set is
\begin{equation}
\mathcal{M}^{\star}_{s,\tau,\ell}
=
\operatorname{TopK}_{K_{\mathrm{top}}}
\left\{
S_{s,\tau,\ell,m}: m\in\mathcal{M}_{\ell}
\right\}.
\label{eq:topk_select}
\end{equation}

This step is executed online on all layers. The owner first queries the router score for each candidate expert, and then augments it with latency and congestion terms derived from the current network state. Subsequently, the top-$K$ selection is applied to obtain
$\mathcal{M}^{\star}_{s,\tau,\ell}$, after which the owner sends hidden activations only to the selected expert sites.

\setlength{\textfloatsep}{10pt plus 1pt minus 1pt}
\setlength{\floatsep}{6pt plus 1pt minus 1pt}
\begin{algorithm}[b]
\caption{StateFlow Execution Policy}
\label{alg:stateflow}
\renewcommand{\algorithmicrequire}{\textbf{Input:}}
\renewcommand{\algorithmicensure}{\textbf{Output:}}
\begin{algorithmic}[1]
\REQUIRE Dialogue $s$, entry site $e_s$, deployment map $\{\nu_{\ell}\}$
\ENSURE Sticky owner $o_s$, routing variables $z$, aggregation sites $a$
\STATE Select owner $o_s$ by \eqref{eq:owner_select}
\FOR{each turn $\tau$ of dialogue $s$}
    \FOR{each layer $\ell$}
        \STATE Compute local hidden state $h_{s,\tau,\ell}$
        \FOR{each expert $m\in\mathcal{M}_{\ell}$}
            \STATE Evaluate $S_{s,\tau,\ell,m}$ by \eqref{eq:expert_score}
        \ENDFOR
        \STATE Select $\mathcal{M}^{\star}_{s,\tau,\ell}$ by \eqref{eq:topk_select}
        \STATE Select $a_{s,\tau,\ell}^{\star}$ by \eqref{eq:agg_select}
        \STATE Dispatch remote branches and execute experts in parallel
        \STATE Aggregate outputs at $a_{s,\tau,\ell}^{\star}$
    \ENDFOR
    \STATE Update ${K}_{s,\ell}^{(\tau)}$
    \STATE Update congestion prices by \eqref{eq:dual_update}
\ENDFOR
\end{algorithmic}
\end{algorithm}

% \vspace{-0.2cm}
\subsection{Path-Aware Aggregation}
% \vspace{-0.1cm}

After determining the active experts, StateFlow selects the aggregation site for merging their outputs. Aggregating at the owner site preserves continuity and avoids an additional return hop, but incurs return delays. Choosing a remote expert site can shorten some return paths, but may increase the cost of sending the merged result back to the owner. 

StateFlow treats aggregation as a path-aware placement decision, allowing the aggregation site to be dynamically selected based on the communication path.
\begin{equation}
a_{s,\tau,\ell}^{\star}
=
\operatorname*{arg\,min}_{\,a\,\in\,\{o_s\}\cup\{v_m:\,m\in\mathcal{M}^{\star}_{s,\tau,\ell}\}}
\Lambda_{s,\tau,\ell}(a).
\label{eq:agg_select}
\end{equation}

StateFlow maintains online congestion prices to prevent persistent overuse of hot sites:
\begin{equation}
\alpha_v(t+1)
=
\left[\alpha_v(t)
+\delta\!\left(\sum_{u\in\mathcal{U}_v(t)}\!\kappa_u(t)-C_v(t)\right)\right]^{\!+},
\label{eq:dual_update}
\end{equation}
where $\delta>0$ is the step size and $[\,\cdot\,]^+=\max\{\,\cdot\,,0\}$.

Combining the above components, StateFlow forms a per-dialogue online control loop. It first selects a sticky owner to anchor the KV state, then performs latency- and congestion-aware expert routing, path-aware aggregation, and congestion-price updates at each turn and MoE layer. In this way, persistent state remains stable while sparse computation adapts to current network and load conditions. Overall, the StateFlow execution policy is summarized in Algorithm~\ref{alg:stateflow}.

\begin{figure}[!t]
    \centering
    \includegraphics[
        width=\columnwidth,
        trim=45 60 420 37, % 左 下 右 上
        clip
    ]{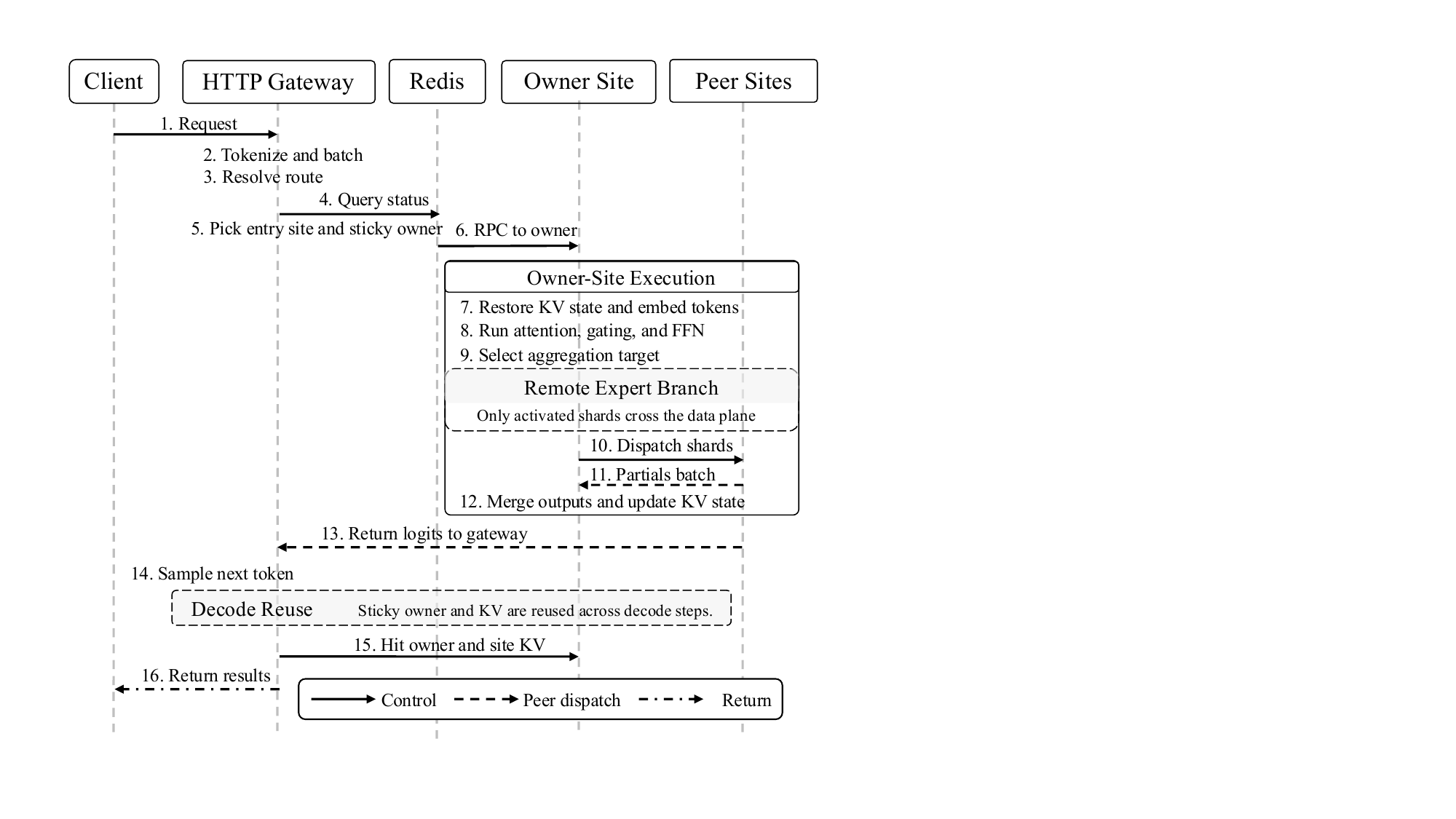}
    \vspace{-0.5cm}
    \caption{StateFlow online serving workflow: owner routing, expert dispatch, path-aware aggregation, and KV state update.}
    % \vspace{-0.1cm}
    \label{fig:online_workflow}
    % \vspace{-2mm}
\end{figure}

% \vspace{-0.1cm}
\subsection{Online Serving Workflow}

StateFlow connects the gateway, controller, sticky owner, and peer
expert sites in an online serving loop. Upon request arrival, the gateway
tokenizes and batches the input, queries the controller, and resolves the
entry site and sticky owner. The owner restores the dialogue KV state,
executes owner-side attention and routing, and dispatches hidden
activations to selected remote experts when needed. Expert outputs are
merged at the selected aggregation site and returned to the owner to
continue layer execution and update the KV state. After the turn
completes, the owner returns logits to the gateway for token sampling
and response generation. During decode, the same dialogue identifier
directly returns to the sticky owner, enabling KV reuse across steps.
The overall serving workflow is summarized in
Fig.~\ref{fig:online_workflow}.

% \vspace{-0.2cm}
\section{Experiments and Results}
\label{sec:exp}
% \vspace{-0.1cm}

\subsection{Experimental Setup}
\label{subsec:testbed}

Experiments are conducted on a single server equipped with eight NVIDIA A100 GPUs. Six GPUs instantiate the serving sites, with one site agent pinned to one GPU and isolated inside one Linux network namespace. The remaining GPUs host the gateway and control-plane processes and are excluded from the serving-site resource budget. The testbed realizes a three-tier hierarchy with four $\mathcal{V}_1$ sites, one $\mathcal{V}_2$ site, and one $\mathcal{V}_3$ site. Expert shards are co-located with serving sites: each $\mathcal{V}_1$ site hosts one shard, while $\mathcal{V}_2$ and $\mathcal{V}_3$ each host two.

The network fabric is implemented with namespace-to-namespace \textit{veth} pairs. The shaped data plane contains four $\mathcal{V}_1$--$\mathcal{V}_2$ $(25\,\text{Gbps},\,5\,\text{ms})$ links, one $\mathcal{V}_2$--$\mathcal{V}_3$ $(80\,\text{Gbps},\,20\,\text{ms})$ link, and six direct $\mathcal{V}_1$--$\mathcal{V}_1$ $(10\,\text{Gbps},\,2\,\text{ms})$ links. Linux \textit{tc} rules are installed on both directions of each data-plane interface, combining token-bucket bandwidth control with \textit{netem} queueing and latency emulation. Inter-site activation dispatch, expert-output return, and aggregation traffic therefore traverse the kernel TCP/IP stack and are subject to actual transmission, bandwidth throttling, and queue buildup. A separate host--namespace control plane is used for orchestration, health checks, and logging, so management traffic does not bypass or perturb the shaped data-plane measurements. Testbed configuration and workload parameters are listed in Table~\ref{tab:system_params}.

\begin{table}[!t]
\centering
\caption{Testbed configuration and workload parameters.}
\vspace{-0.2cm}
\label{tab:system_params}
\scriptsize
\setlength{\tabcolsep}{3pt}
\renewcommand{\arraystretch}{1}
\begin{tabular*}{\columnwidth}{@{\extracolsep{\fill}}ll@{}}
\toprule
\multicolumn{2}{@{}l@{}}{\textit{System}} \\
\midrule
Model & Mixtral-8x7B-Instruct \\
Transformer layers & 32 \\
Experts / MoE layer & 8 (top-$K{=}2$) \\
Weight dtype & FP16 \\
Max KV context & 1536 tokens \\
Max dialogues / site & 8 \\
KV TTL & 120\,s \\
$(\omega_{\mathrm{1}}, \omega_{\mathrm{2}})$ & $(0.5,\,0.5)$ \\
$(\lambda, \beta, \gamma)$ & $(1/3,\ 1/3,\ 1/3)$ \\
\midrule
\multicolumn{2}{@{}l@{}}{\textit{Workload}} \\
\midrule
Stable-serving threshold & success rate $\ge 0.95$ \\
End-to-end turn completion budget & 300\,s \\
Ingress pattern & uniform \\
Capacity sweep & $\{4,8,16,20,24,28,32,36,40\}$ \\
Background load & $1.5$\,req/s, $\leq 8$ in-flight \\
Effective $\mathcal{V}_1$ capacity factor & 0.75 \\
\bottomrule
\vspace{-0.2cm}
\end{tabular*}
\end{table}

\subsection{Results}
\label{subsec:results}

To evaluate the performance of StateFlow, we compare it against three distributed baselines: CommRoute~\cite{xue2025wdmoe}, a communication-aware MoE routing ablation that omits KV state continuity preservation; Llumnix~\cite{sun2024llumnix}, which uses reactive request migration and adaptive deployment; and PReQuaL~\cite{wydrowski2024load}, which applies latency and queue load balancing without persistent KV state. For fairness, all baselines use the same model, topology, expert placement, GPU allocation, and tc-netem link profiles. All distributed schemes attain the same MMLU subset accuracy of 56.25\%, compared with 58.04\% for a centralized cloud baseline, confirming that distributed execution preserves model quality. The following experiments focus on latency and concurrency.

\subsubsection{Stable Dialogue Concurrency}

\begin{figure}[!t]
    \centering
    \begin{minipage}[t]{0.49\columnwidth}
        \centering
        \includegraphics[width=\linewidth]{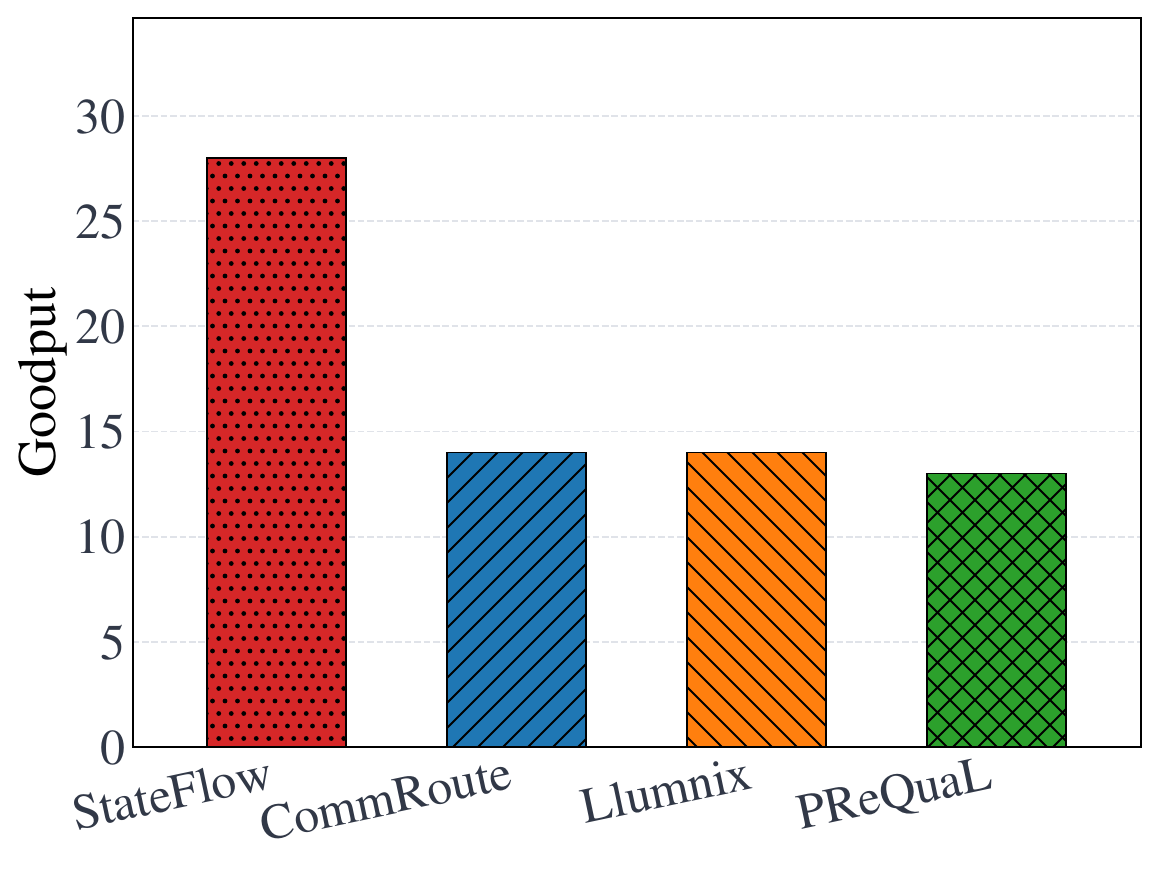}
    \end{minipage}\hfill
    \begin{minipage}[t]{0.5\columnwidth}
        \centering
        \includegraphics[width=\linewidth]{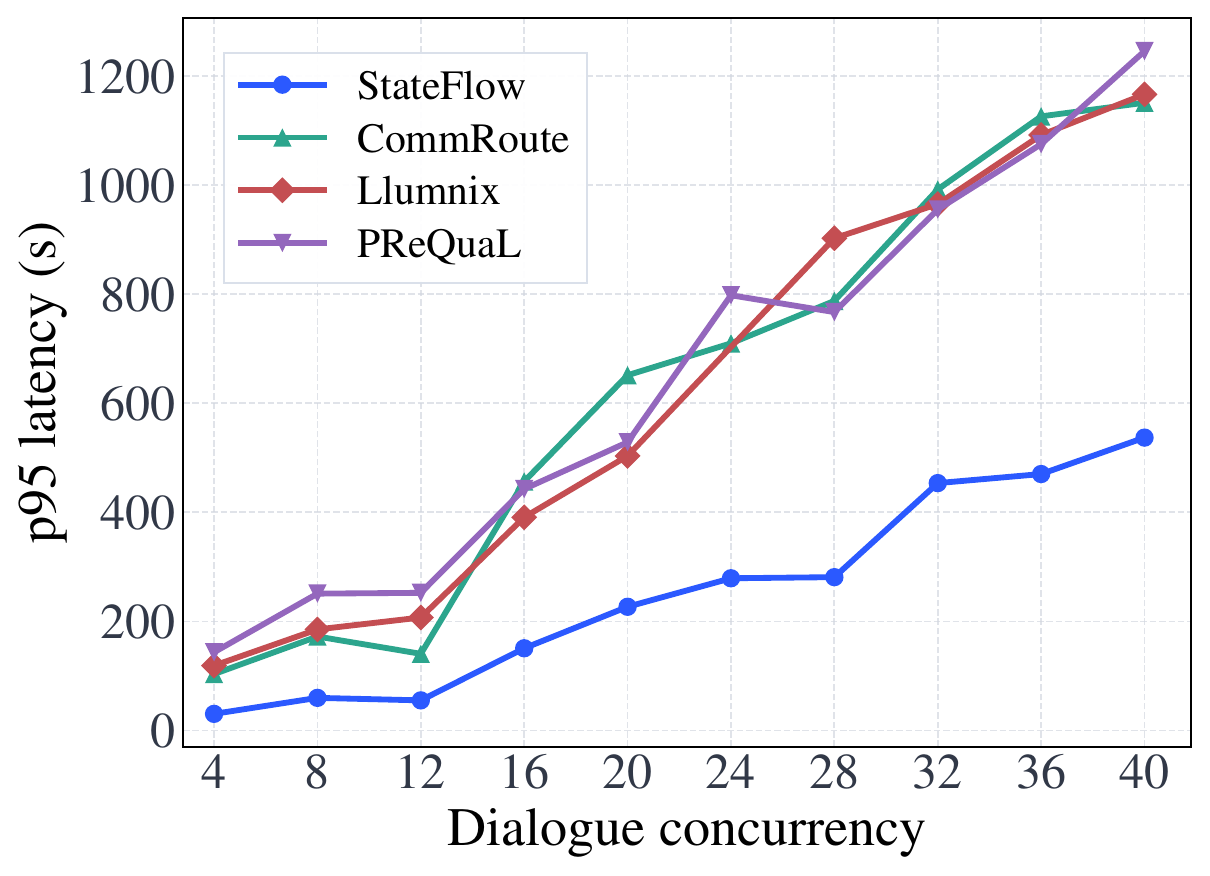}
    \end{minipage}
    \vspace{-0.8cm}
    \caption{Stable dialogue concurrency under uniform load: goodput (left) and p95 turn-level completion latency (right) versus concurrent dialogues~$W$.}
    \label{fig:capacity_mean_p95}
    \vspace{-0.2cm}
\end{figure}

To determine the concurrency level each method can sustain without violating the turn-level latency budget, we sweep the offered dialogues $W$ from 4 to 40. Fig.~\ref{fig:capacity_mean_p95} reports goodput and p95 turn-level completion latency across this range. StateFlow sustains the latency budget up to $W{=}28$. All baselines meet the budget below $W{=}13$, but beyond this point they exhibit a steep p95 latency rise and an abrupt goodput collapse, while StateFlow's curves remain nearly flat over the same range. Therefore, StateFlow sustains more than $2\times$ the stable concurrency of the strongest baseline before compute capacity itself becomes the binding constraint.

Table~\ref{tab:capacity_w24} details the latency distribution and throughput at StateFlow's stable concurrency boundary $W{=}28$. StateFlow is the only scheme that remains within the 300\,s p95 budget, achieving a p95 of 281\,s and a throughput of 0.107\,rps. Under the same load, CommRoute reaches 787\,s p95 with only 0.036\,rps, corresponding to a 2.9$\times$ higher mean latency and a nearly 3.0$\times$ lower throughput. Llumnix and PReQuaL perform similarly, with p95 latencies of 902\,s and 767\,s and throughputs of 0.033 and 0.036\,rps respectively. StateFlow also yields the tightest latency distribution, with a standard deviation of 22\,s against 55\,s for CommRoute and 72\,s for Llumnix. The combination of low tail latency and tight variance indicates that StateFlow maintains predictable per-turn performance even near its capacity boundary.

\begin{figure}[!t]
    \centering
    \begin{minipage}[t]{0.51\columnwidth}
        \centering
        \includegraphics[width=\linewidth]{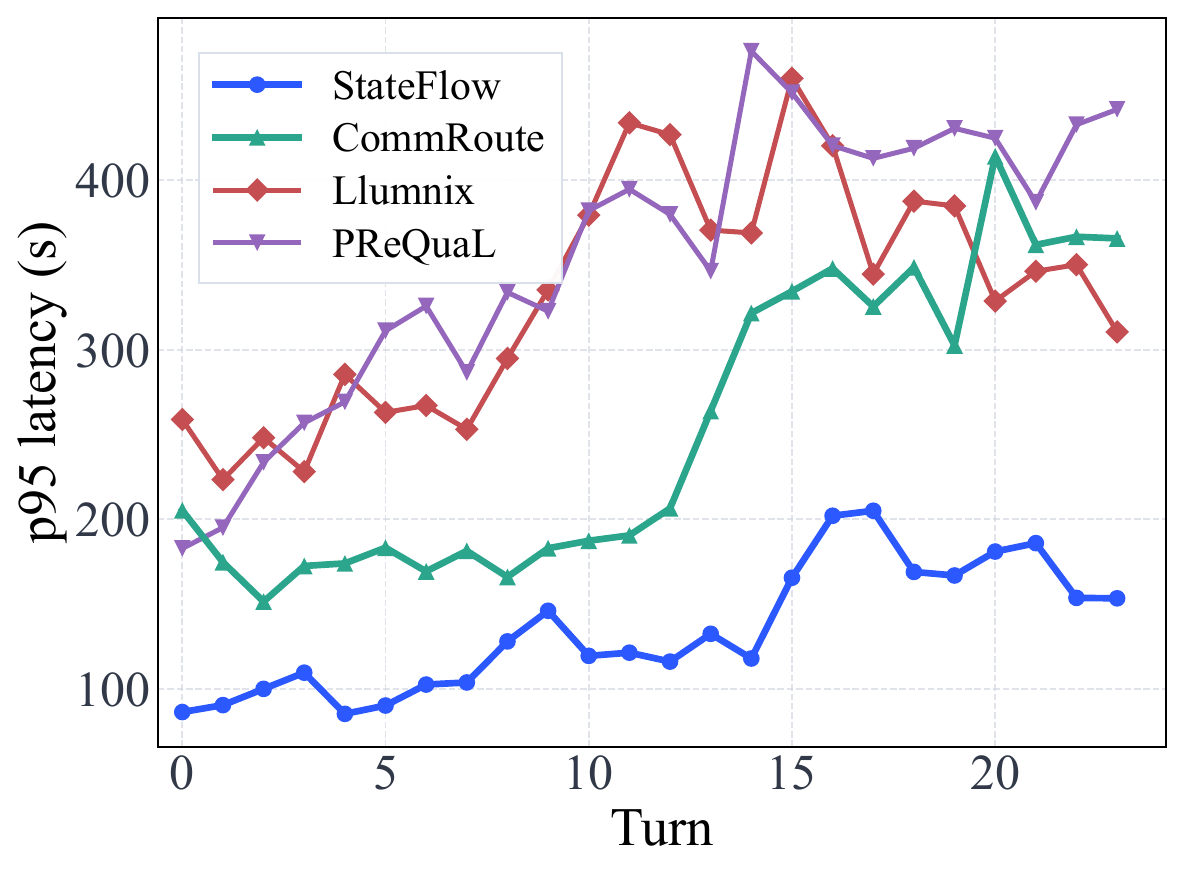}
    \end{minipage}\hfill
    \begin{minipage}[t]{0.48\columnwidth}
        \centering
        \includegraphics[width=\linewidth]{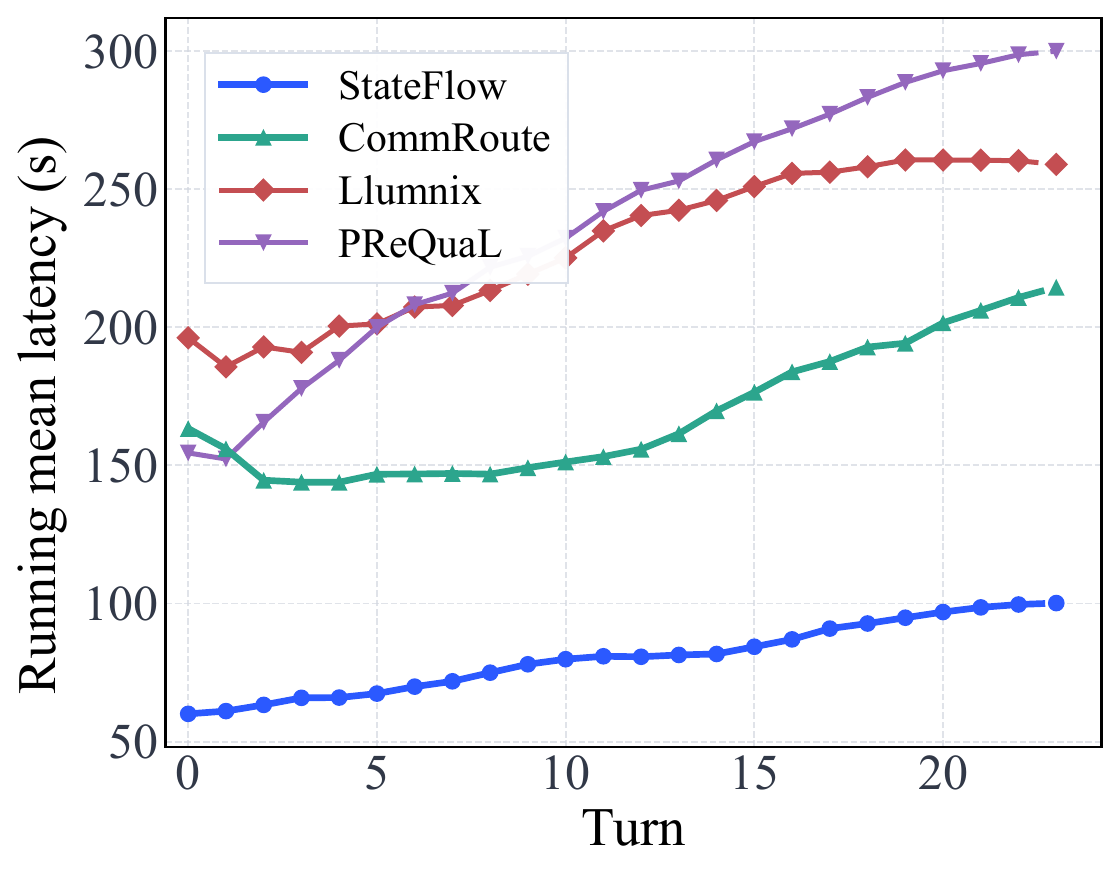}
    \end{minipage}
    \vspace{-0.7cm}
    \caption{Multi-turn inference dynamics over 24 consecutive turns across 6 concurrent dialogues: per-turn p95 completion latency across concurrent dialogues (left) and running mean turn completion latency(right).} 
    \vspace{-0.2cm}
    \label{fig:multiturn_latency}
    % \vspace{-2mm}
\end{figure}

\begin{table}[!t]
\centering
\caption{Turn-level completion latency distribution and throughput at the StateFlow stable concurrency boundary $W{=}28$.}
\vspace{-0.1cm}
\label{tab:capacity_w24}
\renewcommand{\arraystretch}{1.08}
\setlength{\tabcolsep}{2pt}
\footnotesize
\begin{tabular*}{\columnwidth}{@{\extracolsep{\fill}}lcccccc@{}}
\toprule
\textbf{Baseline} & \textbf{Mean (s)} & \textbf{P50 (s)} & \textbf{P95 (s)} & \textbf{P99 (s)} & \textbf{Std. (s)} & \textbf{Thr. (rps)} \\
\midrule
StateFlow & \textbf{244} & \textbf{242} & \textbf{281} & \textbf{294} & \textbf{22} & \textbf{0.107} \\
CommRoute & 714 & 724 & 787 & 819 & 55 & 0.036 \\
Llumnix & 803 & 816 & 902 & 953 & 72 & 0.033 \\
PReQuaL & 720 & 726 & 767 & 797 & 41 & 0.036 \\
\bottomrule
\end{tabular*}
\vspace{-0.1cm}
\end{table}

\subsubsection{Multi-Turn Latency Dynamics}

To verify that per-turn latency remains stable as KV state accumulates across turns, we run 6 concurrent dialogues of 24 turns each. As shown in Fig.~\ref{fig:multiturn_latency}, StateFlow attains the lowest latency throughout, with a mean of 101.8\,s and a p95 of 170.4\,s, yielding a 52.5\% reduction in mean and 53.0\% reduction in p95 against CommRoute. Llumnix and PReQuaL trail CommRoute throughout the trace. StateFlow's advantage widens with dialogue depth, completing the final turn in 111.9\,s compared with 300.3\,s for CommRoute. This widening gap confirms that the benefit of sticky owner selection compounds over turns, as each successive turn reuses the anchored KV state instead of incurring repeated cross-site transfers.

Table~\ref{tab:remote_dispatch_kvhit} decomposes per-turn execution into remote expert dispatch latency and KV hit rate. The per-turn KV hit rate is defined as the fraction of layer-0 attention lookups at the owner site that retrieve valid KV states from the prior turn. All four methods achieve KV hit rates above 94\%, confirming that KV state is effectively reused across turns in each scheme. Given comparable KV reuse, the performance separation is driven by remote dispatch latency. At $W{=}28$, StateFlow reduces remote dispatch latency to 468.5\,ms, whereas CommRoute, Llumnix, and PReQuaL incur 1408.7\,ms, 1835.4\,ms, and 1466.6\,ms respectively. This 3$\times$ reduction accounts for StateFlow's capacity advantage in Fig.~\ref{fig:capacity_mean_p95}, which emerges in the mid-to-high concurrency regime where dispatch contention becomes the dominant cost for methods that lack latency-aware routing.

\begin{table}[!t]
\centering
\caption{Remote expert dispatch latency and KV hit rate across schemes at the capacity boundary $W{=}28$.}
\label{tab:remote_dispatch_kvhit}
\renewcommand{\arraystretch}{1.08}
\setlength{\tabcolsep}{3pt}
\footnotesize
\begin{tabular*}{\columnwidth}{@{\extracolsep{\fill}}lcc@{}}
\toprule
\textbf{Baseline} & \textbf{Remote Dispatch Latency (ms)} & \textbf{Per-turn KV hit rate (\%)} \\
\midrule
StateFlow & \textbf{468.5} & \textbf{96.7} \\
CommRoute & 1408.7 & 96.6 \\
Llumnix   & 1835.4 & 94.2 \\
PReQuaL   & 1466.6 & 94.7 \\
\bottomrule
\end{tabular*}
\vspace{-2mm}
\end{table}

\begin{figure}[!t]
    \centering
    \begin{minipage}[t]{0.51\columnwidth}
        \centering
        \includegraphics[width=\linewidth]{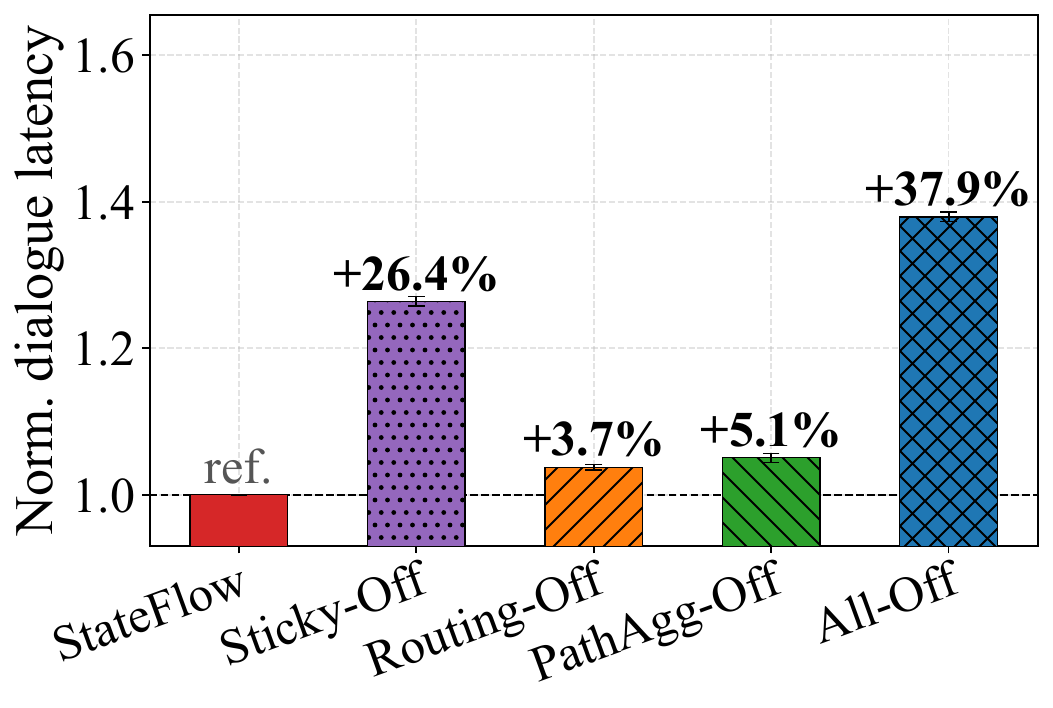}
    \end{minipage}\hfill
    \begin{minipage}[t]{0.48\columnwidth}
        \centering
        \includegraphics[width=\linewidth]{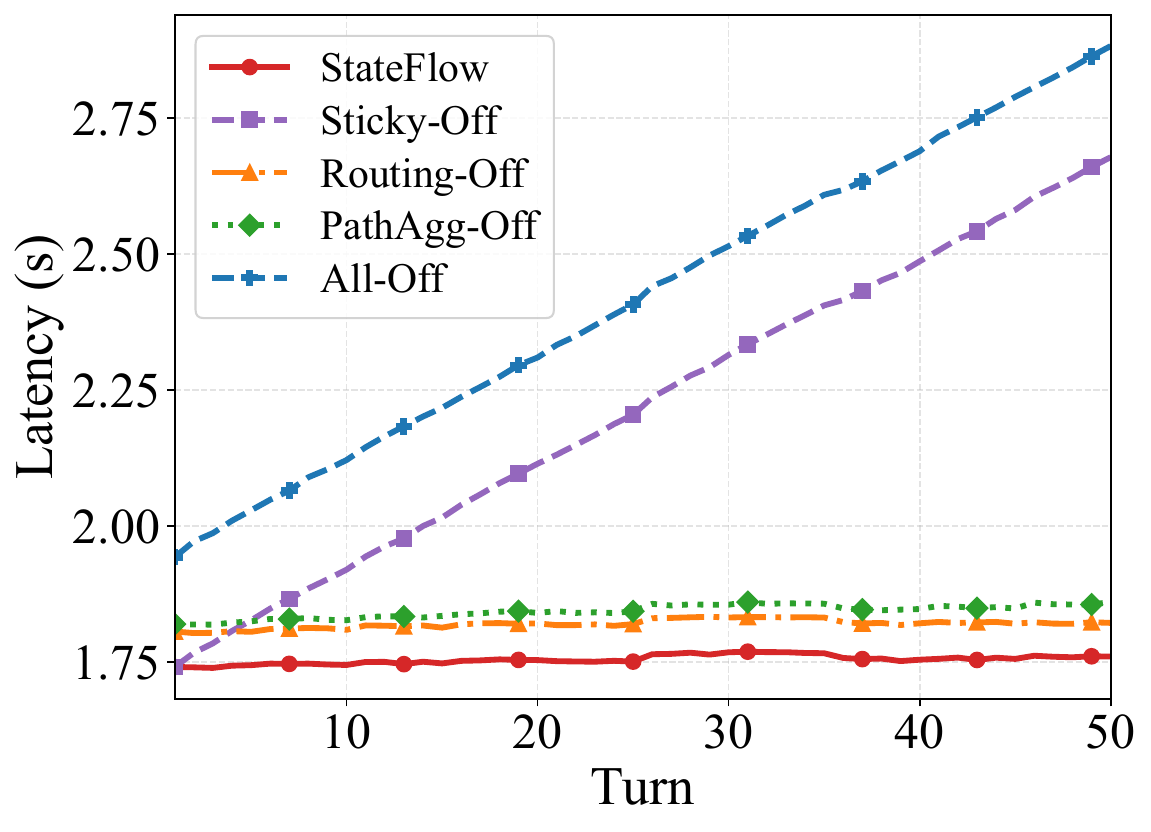}
    \end{minipage}
    \vspace{-0.7cm}
    \caption{Ablation of StateFlow over 50-turn dialogues. The bar chart reports normalized total dialogue latency, while the curve plot shows per-turn latency.}
    \vspace{-0.1cm}
    \label{fig:ablation}
\end{figure}

\subsubsection{Ablation study}
To quantify the contribution of each component to dialogue serving, Fig.~\ref{fig:ablation} reports an ablation experiment over 50-turn dialogues. \textit{Sticky-Off} causes the largest degradation, increasing mean dialogue latency by \textbf{26.4}\%. Without a sticky owner, the accumulated KV context must be transferred to a new site across turns, causing the penalty to grow with dialogue depth, with latency increasing from $1.75$\,s to $2.65$\,s. \textit{Routing-Off} and \textit{PathAgg-Off} lead to smaller increases of \textbf{3.7}\% and \textbf{5.1}\%, respectively. Their penalties remain nearly flat across turns because these two mainly reduce per-turn expert dispatch and aggregation costs, with limited impact on cross-turn state movement. When all components are removed, mean latency increases by \textbf{37.9}\%, and the tail latency rises by \textbf{64.6}\%, showing a compounding effect among state migration, expert dispatch, and aggregation overhead. These results demonstrate that the three components target complementary
sources of inefficiency. Sticky ownership provides the dominant gain for long dialogues, while latency-aware routing and path-aware aggregation deliver consistent per-turn improvements.

\section{Conclusion}

This paper presented \textit{StateFlow}, a multi-turn distributed inference policy for MoE serving over 6G edge--cloud networks. 
StateFlow decouples persistent KV state from transient sparse
computation by anchoring ownership and KV state near the access
edge for continuity, while selectively spilling sparse expert execution
across the hierarchy to exploit available capacity. Experiments on a real-world testbed with kernel-level network emulation show that this policy sustains over $2\times$ higher dialogue concurrency, reduces p95 latency by $53.0\%$, and cuts remote dispatch overhead by 66.7\% relative to the strongest baseline, while preserving the highest KV reuse rate.

\section*{Acknowledgments}
\small{This work was supported by the UK Engineering and Physical Sciences Research Council (EPSRC) grant EP/Y037243/1, EP/X04047X/2 for the TITAN Telecoms Hub and the Federated Telecoms Hubs, and grant EP/Y036514/1 for the JOINER project.
}

\vspace{-0.1cm}
\bibliographystyle{IEEEtran}
\bibliography{references}

\end{document}